# Silicon Nitride Waveguides for Plasmon Optical Trapping and Sensing Applications

Qiancheng Zhao[1], Caner Guclu[1], Yuewang Huang[1], Filippo Capolino[1] and Ozdal Boyraz*[1]
[1]Department of Electrical Engineering and Computer Science, University of California-Irvine, Irvine, California, USA, 92697
*Email: oboyraz@uci.edu

## ABSTRACT

We demonstrate a silicon nitride trench waveguide deposited with bowtie antennas for plasmonic enhanced optical trapping. The sub-micron silicon nitride trench waveguides were fabricated with conventional optical lithography in a low cost manner. The waveguides embrace not only low propagation loss and high nonlinearity, but also the inborn merits of combining micro-fluidic channel and waveguide together. Analyte contained in the trapezoidal trench channel can interact with the evanescent field from the waveguide beneath. The evanescent field can be further enhanced by plasmonic nanostructures. With the help of gold nano bowtie antennas, the studied waveguide shows outstanding trapping capability on 10 nm polystyrene nanoparticles. We show that the bowtie antennas can lead to 60-fold enhancement of electric field in the antenna gap. The optical trapping force on a nanoparticle is boosted by three orders of magnitude. A strong tendency shows the nanoparticle is likely to move to the high field strength region, exhibiting the trapping capability of the antenna. Gradient force in vertical direction is calculation by using a point-like dipole assumption, and the analytical solution matches the full-wave simulation well. The investigation indicates that nanostructure patterned silicon nitride trench waveguide is suitable for optical trapping and nanoparticle sensing applications.

**Keywords:** silicon nitride trench waveguide, optical trapping, micro-fluidic channel, bowtie antenna, gradient force.

## 1. INTRODUCTION

A novel approach of silicon nitride trench waveguide has been put forward recently with the ability to combine microfluidic channel and waveguide together [1], [2]. Analyte, contained in the trapezoidal trench, interacts with the field tailing away from the trench waveguide, which can be enhanced by a plasmonic bowtie antenna. Plasmon enhanced field is widely used in optical tweezing such as nanoparticle trapping [3], molecules sensing [4] and bio cell control [5]. Here we investigate the optomechanical properties of a silicon nitride trench waveguide enhanced by gold bowtie antennas. The bowtie antenna leads to 60-fold enhancement of electric field in the antenna gap. With the help of scattering by the antenna, the optical trapping force on a 10 nm radius polystyrene nanoparticle is boosted by 3 orders of magnitude. The strong tendency of a nanoparticle moving to the high field intensity region is obtained, exhibiting the trapping capability of the antenna. The device is promising for particle sensing and sorting.

## 2. WAVEGUIDE AND BOWTIE ANTENNA DESIGN

### 2.1 Waveguide design and fabrication

Recently we demonstrated a method to fabricate sub-micron silicon nitride trench waveguide using conventional lithography, which has a low propagation loss of 0.8 dB/cm and high nonlinearity $n_2 = 1.39 \times 10^{-19}$ m$^2$/W [1], [2]. Fabrication of the waveguide is rather straight forward and does not require E-beam lithography to achieve sub-micron waveguide dimension, and merely relies of optical lithography followed by anisotropic potassium hydroxide etching. The anisotropic potassium hydroxide (KOH) etching on <100> silicon wafer carves a trapezoidal or triangular trench with an angle of 54.7° with respect to the substrate surface, because the etching rate on the <111> direction is extremely slow. The shape of the waveguide is determined by etching time, opening window and silicon nitride deposition thickness. For example, if the etching opening window is smaller than $2 \times \sqrt{2}/2 \times H$, where $H$ is the etching depth on the silicon substrate, a V-groove will be carved out, and the fabricated waveguide will be in triangle shape. On the contrary, if the etching width is larger than that value, KOH will carve isosceles trapezoidal trench. The width of the lower edge of the waveguide can be estimated as $W_{bottom} = W_{open} - 2 \times \sqrt{2}/2 \times H$, where $W_{bottom}$ is the width of the lower edge of the waveguide, and $W_{open}$ is the width of the opening window on photo masks. Low-pressure-chemical-vapor-deposition (LPCVD) was used to deposit a layer of $T = 725$ nm silicon nitride.

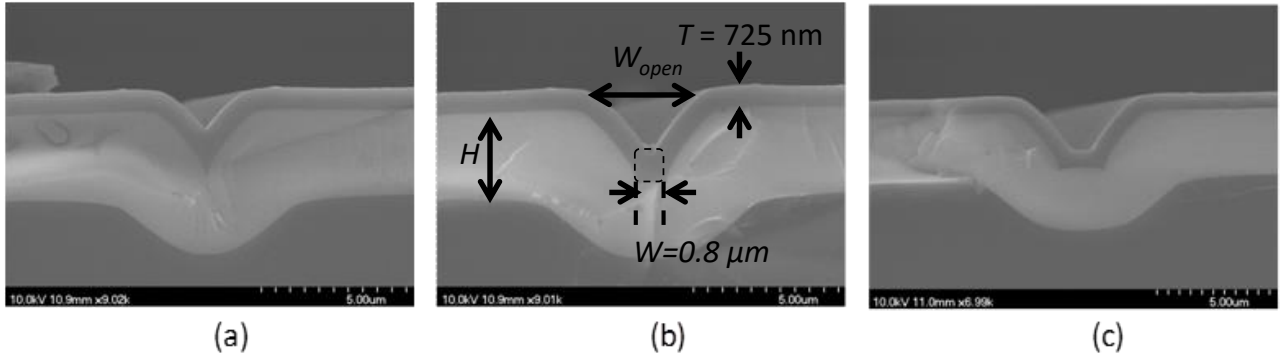

FIG. 1 Fabricated waveguides (a) $W_{open} = 4$ μm, (b) $W_{open} = 5$ μm, and (c) $W_{open} = 6$ μm.

The variation of the waveguide geometry affects the mode distribution and electrical field filed orientation. Wide waveguides favor TE mode while narrow waveguides support TM mode better. The experimental characterized propagation losses in different trench waveguide are available in [6]. It is worth noting that TE mode is preferred for plasmonic sensing applications, and this will be discussed later in antenna design. The dispersion curve for the TE mode is plotted as a function of waveguide bottom width.

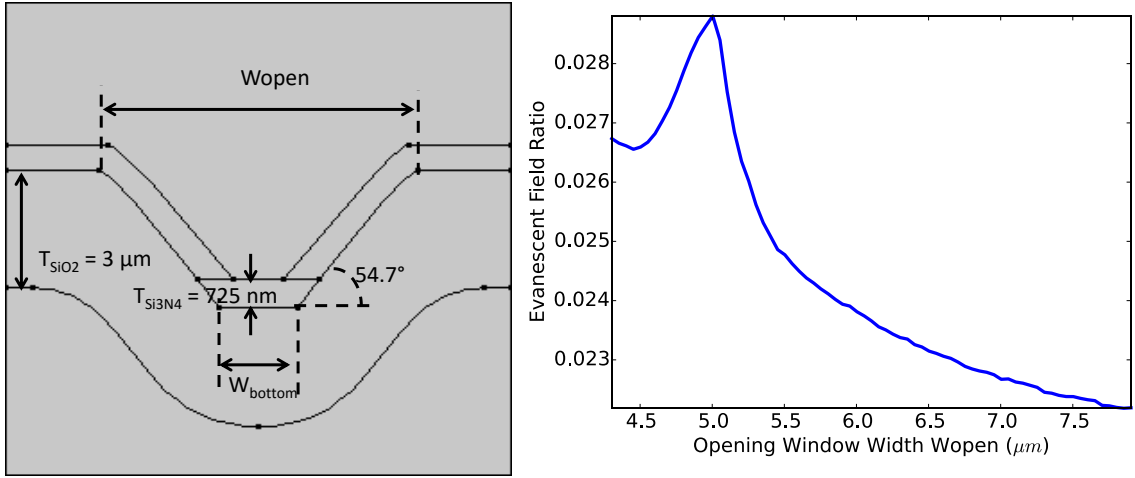

FIG. 2. (a) Structure of a silicon nitride trench waveguide. (b) TE mode dispersion curve as function of waveguide bottom width. (c) $E_{11}^x$ mode profile when $W_{bottom} = 1.85$ μm. (d) $E_{21}^x$ mode profile when $W_{bottom} = 4.25$ μm. (e) $E_{31}^x$ mode profile when $W_{bottom} = 5.65$ μm.

## 2.2 Antenna design

The optical gradient forces experienced by nanoparticles are typically very weak (some femtonewtons or less), because the dipolar polarizability scales with the third power of the particle size. The volume-scaling of the maximum trapping force was evaluated explicitly in ref [7] for polystyrene spheres, showing a decrease of three orders of magnitude in the maximum trapping force as the sphere radius decreased from 100 to 10 nm. Therefore, to confine nanoparticles against the destabilizing effects of thermal fluctuations, a significantly higher optical power is required. The plasmonic nature of metal nanostructure can enhance the electrical field intensity, so that stable trapping can be achieved at a much lower power.

In recent years, bowtie antenna arrays for particle manipulation in objective lens configuration have been extensively studied [8]. When incident light is linearly polarized across the gap of a bowtie antenna, capacitive effects lead to a confined and intense electric field spot in antenna near field particularly within the gap region [4]. However, to the author's knowledge, the use of a waveguide combined with a bowtie antenna for optical manipulation has not been demonstrated yet. To facilitate coupling from a waveguide mode to the antennas, TE mode is preferred due to its significant tangential electric field with respect to the top surface of the waveguide. The waveguide that is studied here is designed to be 0.5 μm wide at the top, 1.15 μm wide at bottom and has a height of 725 nm in favor of TE mode. A bowtie antenna, made of two equilateral triangle gold patches separated by a gap along the x direction, can be deposited using nanosphere lithography [9]. By manipulating the side length of the antennas shown in FIG. 3(b), resonance can be altered. Although smaller gap leads to stronger field, the gap of bow-tie antenna is set to be 30 nm for the sake of particle size and conforming to fabrication limitations. The metal thickness of the antennas is set to be 20 nm. In the simulation of 10 mW launched power at 1550 nm wavelength, the peak intensity, reported in FIG. 3(a), in the antenna gap (10 nm above the waveguide) is 60 times larger than that in absence of the antenna, at the same location.

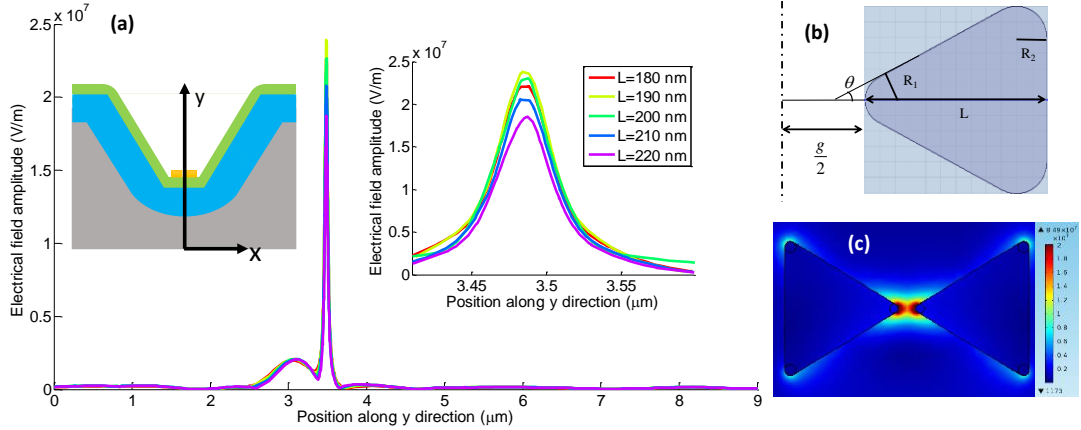

FIG. 3 (a) Electrical field profile along the y direction. The left inset shows the cutline in the structure. The right inset shows the maximum electric field for different antenna lengths. (b) Half structure of a bowtie antenna. $g = 30$ nm in our case.

## 3. OPTICAL FORCE SIMULATION

### 3.1 Optical force simulation

A polystyrene particle with 10 nm radius immersed in water is placed in the gap of bowtie antennas. Optical force is calculated by integration of Maxwell stress tensor over the particle surface. As shown in FIG. 4(a), the vertical optical force magnitude increases first and then decreases as the nanoparticle is elevated away from the waveguide. Maximum force occurs when the particle bottom surface is 14 nm away from the waveguide, at which position the trapping force is boosted by 3 orders of magnitude compared to that occurring without antenna enhancement. The peak vertical optical force is estimated to be $6.8 \times 10^8$ times larger than the net gravitational and buoyancy force.

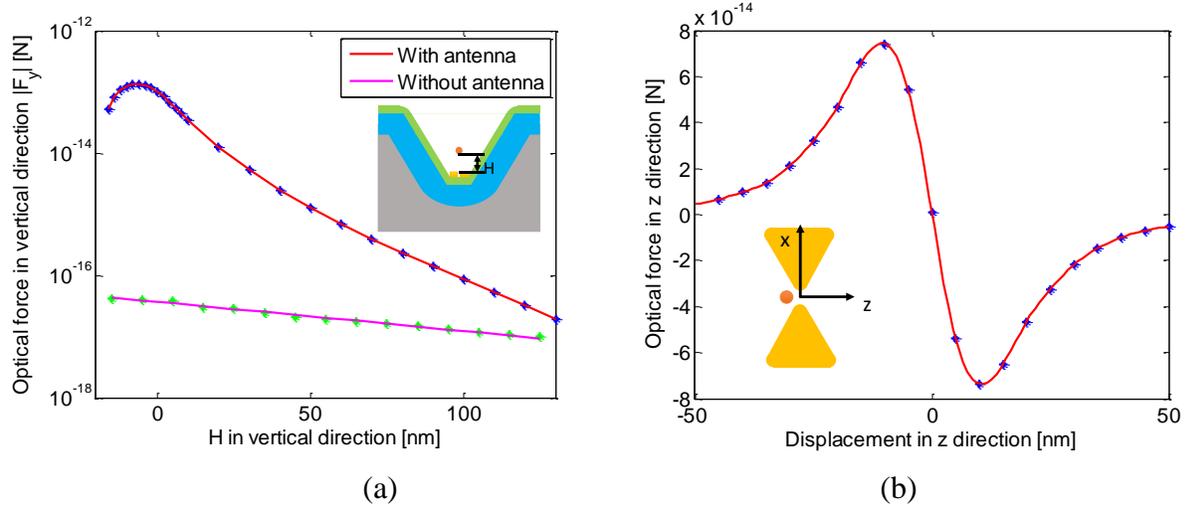

(a) (b)

FIG. 4. (a) Optical trapping force in vertical direction (y axis) vs. vertical distance from the particle bottom surface to the antenna top surface. The force is enhanced by bowtie antenna compared to non-antenna case. (b) Optical trapping force in longitudinal direction (z axis) vs. displacement from the antenna center. The particle bottom surface is 4 nm below the antenna top surface.

FIG. 4(b) illustrates that the longitudinal optical force (z component) tends to attract the nanoparticle back when it is aberrant to the antenna center. The simulation was carried out when the particle is partially immersed in the antenna gap ($H = -6$ nm, FIG. 4(a)). The force is positive when the particle is in negative position, and vice versa. The maximum force occurs when the particle is 10 nm away from the center. The trapped particle can be optically detected by nonlinear optical response such as two-photon fluorescence (TPF) and second harmonic generation [10].

### 3.2 Gradient force verification

The optical trapping force was generated in a full-wave simulator (COMSOL Multiphysics). To verify the simulation, we chose the *y* component of the force and calculate it analytically. Suppose a particle is enough far away from the antenna, so that it does not interfere with the antenna scattering pattern. Then the electrical field of the nanoparticle and antenna system can solely be represented by the antenna field distribution. Furthermore, the particle size is small enough ($R = 10$ nm) compared to wavelength (1550 nm), so it can be viewed as a point. The optical force experienced by a nanoparticle is then modeled in a dipole model. For a sphere of radius *a* and relative permittivity of $\varepsilon_r$, the point-like particle polarizability $\alpha_0$ is given by the Clausius-Mossotti relation [11], [12],

$$\alpha_0 = 4\pi n_2^2 \varepsilon_0 a^3 \frac{m^2 - 1}{m^2 + 2} \tag{1}$$

where $n_2$ is the surrounding medium refractive index, $m = n_1/n_2$ is the ratio of the refractive index of the particle and the medium, $\varepsilon_0$ is the vacuum permittivity. If taken the reaction of a finite-size dipole to the scattered field at its own location, the Clausius-Mossotti relation can be corrected as follows [13]:

$$\alpha = \frac{\alpha_0}{1 - \frac{ik^3 \alpha_0}{6\pi\varepsilon_0}} \tag{2}$$

The explicit form of the time-averaged force acting on such a dipole is:

$$\langle \mathbf{F} \rangle = \frac{1}{4}\operatorname{Re}(\alpha)\nabla|\mathbf{E}|^2 + \frac{\sigma}{2c}\operatorname{Re}(\mathbf{E}\times\mathbf{H}^*) + \frac{\sigma c \varepsilon_0}{4\omega i}\nabla\times\mathbf{E}\times\mathbf{E}^* \tag{3}$$

We scrutinize the region of interest (ROI) where the polystyrene sphere is lifted from 10 nm to 140 nm above the antenna surface. Below this region, the particle will strongly interact with the antenna radiation pattern; while above this region, the force is too small to be considered. The ROI region is between 3.505 μm and 3.625 μm in *y* axis. The simulation was performed with antenna structure parameters $L = 200$ nm and $g = 30$ nm. Due to the electric field enhancement of the antenna, the electric filed is dominated by its *x* component. To calculate the optical force in *y* direction, only the *x* component of electric field and *z* component of the magnetic field are considered. Thereby, Eq. (4) can be written in the following way in favor of $F_z$.

$$\langle F_z \rangle = \frac{1}{4}\text{Re}(\alpha)\frac{\partial}{\partial z}|E_x|^2 + \frac{\sigma}{2c}E_x \cdot H_z - \frac{\sigma c \varepsilon_0}{4\omega i}\frac{\partial}{\partial z}|E_x|^2 \quad (4)$$

Rayleigh scattering cross section scales with $a^6$ when the particle size is smaller than the wavelength [14]. Consider $a = 10$ nm in our case, the contribution of the last two terms are negligible. With the help of extracting the electric and magnetic field in the region of interest, we calculated the optical trapping force in *y* direction, and compared the results generated by COMSOL using Maxwell Stress Tensor (MST) method, shown in FIG. 5.

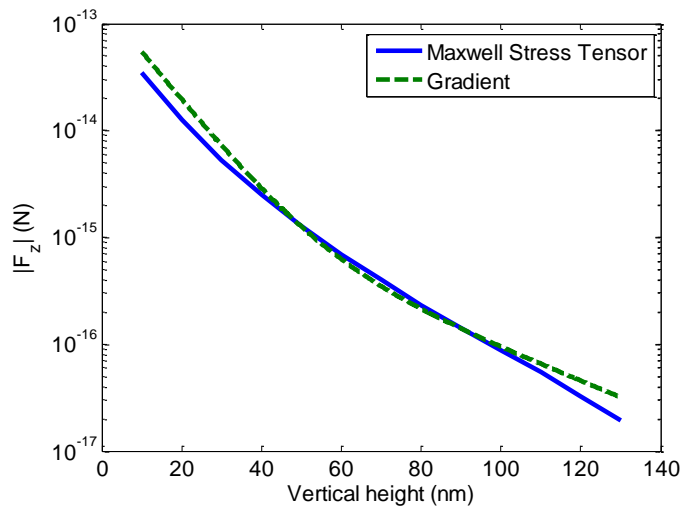

FIG. 5. Comparison of optical gradient force (green) and force by Maxwell stress tensor (blue).

The calculated optical gradient force matches the calculated force well. Both are at the same order of magnitude at all locations, and the trends of the forces remain the same. The mismatches at the beginning of the figure when particle is close to the antenna may come from antenna-nanoparticle interaction. When the particle is far away from the antenna, $E_x$ becomes less dominant, so $F_z$ calculated by only $E_x$ and $H_y$ will have a mismatch with the full simulation. But in total, the mismatches are small.

## 4. CONCLUSION

In conclusion, we demonstrate the optomechanics property of a bowtie antenna deposited on silicon nitride trench waveguide. The plasmon enhanced waveguides exhibit extraordinary trapping capability of polystyrene nanoparticles. The simulated force is boosted by 3 orders of magnitude with the help of bowtie antenna. The force is also analytically studied and its analytical solution matches the results from full-wave simulation. The investigation indicates that nanostructure patterned silicon nitride waveguide is suitable for optical trapping and nanoparticle sensing applications.

## 5. ACEKNOWLEDGMENTS

This work is supported by NSF ECCS 1449397 SNM grant.